\documentclass[11pt]{article}
\usepackage[margin=1in]{geometry}
\usepackage{amsmath,amssymb}
\usepackage{graphicx}
\usepackage[round]{natbib}
\usepackage[colorlinks=true,linkcolor=blue,citecolor=blue,urlcolor=blue]{hyperref}
\title{Fragmentation Dynamics of Pristine Interstellar Comets:\\
An Exploratory Multi-Physics Simulation Study}

\author{Behrooz Karamiqucham\\[1ex]
\normalsize Department of Physics \& Astronomy, College of Charleston,
Charleston, SC~29424, USA\\
\normalsize \texttt{b.karamiqucham@unswalumni.com};
\texttt{karamiquchamb@cofc.edu}}

\begin{document}
\maketitle

\begin{abstract}
\noindent
We present an exploratory numerical model for the thermal evolution and fragmentation of pristine interstellar comets on a first passage through the inner Solar System, applied over a grid of perihelion distances ($q=0.25$--$1.5$~AU) and tensile strengths ($\sigma_t=50$--$500$~Pa). A single nucleus ($M_0=2\times10^{12}$~kg, $R_0\approx1.06$~km, dust/ice $=1$, ices 35\% CO, 30\% CO$_2$, 15\% CH$_4$, 20\% H$_2$O) follows a hyperbolic trajectory from an initially 30~K state, with heat conduction, energy-balanced multi-species sublimation, dust lifting and lag-mantle growth, and a subsurface gas-pressure failure criterion, so that fragmentation is emergent rather than prescribed. In the primary case ($q=1$~AU, $\sigma_t=100$~Pa) splitting begins at the 3~AU starting distance, once a sub-millimeter lag deposit partially confines the warming CO front; 63 binary splittings follow, with a self-limited plateau at 53 bodies before perihelion and the 64-body tracking cap reached near perihelion. Sublimation is energy-limited: total mass loss is 1.3\%, volatile depletions are at most a few per cent, and most of the mass survives as a fragment swarm; the mass budget closes to machine precision. Across the grid, mass loss (0.8--1.7\%) depends weakly on $q$ and $\sigma_t$ but strongly on composition, falling to 0.1\% for depleted, 67P-like ices. Yet every case, including the depleted class, disaggregates, whereas Jupiter-family comets of similar composition survive repeated perihelia, so the sealing prescription over-fragments. We regard the composition ranking, the insensitivity to $\sigma_t$, percent-level mass loss, and an onset beyond 3~AU as robust, and defer absolute survival predictions to calibration against well-observed comets. Mantle insulation suppresses outbound activity, producing an inbound--outbound asymmetry.
\end{abstract}

\medskip
\noindent\textbf{Keywords:} comets: individual (3I/ATLAS) ---
methods: numerical --- minor planets, asteroids: general ---
planets and satellites: formation --- interstellar medium: abundances

\section{Introduction}

The discovery of interstellar objects traversing the inner Solar System
provides unprecedented opportunities to study the products of planet
formation in other stellar systems.  Following the detection of
1I/\textquoteleft Oumuamua in 2017 \citep{meech17}, the identification
of the second known interstellar visitor, 2I/Borisov in 2019
\citep{guzik20}, demonstrated the existence of active cometary bodies
beyond our Solar System.  The recent discovery of 3I/ATLAS (C/2025 N1)
as the third confirmed interstellar comet has generated significant interest; the comet passed perihelion at 1.36~AU on October 29, 2025 \citep{seligman25} and remained active and extensively observed afterwards \citep{cordiner26, opitom26}; post-perihelion Hubble imaging detected a single nucleus of effective radius $1.3\pm0.2$~km \citep{hui26}.  This survival is
particularly noteworthy given the extreme thermal shock experienced by
interstellar objects transitioning from equilibrium temperatures of
$\approx$30~K in interstellar space to $>$200~K near perihelion,
raising fundamental questions about the structural properties and
compositional diversity of cometary material formed in extrasolar systems.  Recent isotopic measurements of 3I/ATLAS
from multiple independent studies \citep{roth26, opitom26, salazar26,
cordiner26} provide strong evidence that 3I formed in an extremely cold
environment with temperatures $<$30~K, consistent with either the outer
regions of a protoplanetary disc or within an interstellar molecular
cloud.  These measurements support the pristine interstellar comet
scenario explored in our simulations and suggest that 3I retained
primordial volatile abundances despite its survival through perihelion.

The structural resilience of 3I/ATLAS contrasts with expectations based
on the fragmentation behavior of some Solar System comets and the
partial disruption observed for 2I/Borisov, which experienced minor
fragmentation approximately three months after its December 2019
perihelion passage \citep{jewitt20}.  This diversity in
fragmentation outcomes---from 1I/\textquoteleft Oumuamua's apparent
complete structural integrity despite unusual non-gravitational
accelerations \citep{micheli18}, to 2I/Borisov's minor disruption, to
3I/ATLAS's survival---suggests that interstellar comets span a range of
material properties, thermal histories, and compositional states.
Understanding this parameter space is crucial for constraining the
formation conditions and evolutionary processing of planetesimals in
extrasolar systems.

Comet fragmentation represents a fundamental process in the dynamical
evolution of small bodies, driven by multiple mechanisms including
thermal stress, rotational instability, tidal disruption, and
collisional cascades \citep{boehnhardt04, sekanina19}.  For interstellar
comets experiencing their first stellar approach after potentially
billions of years in cold interstellar space, the thermal shock is
particularly severe and represents an extreme test of primordial
structural integrity.  Numerical modeling of such events requires
incorporating detailed physics of volatile sublimation, thermal
evolution, and structural failure under rapidly changing thermal and
mechanical stress conditions.

Previous modeling efforts have focused primarily on Solar System comets
with water-dominated compositions and thermal histories involving
multiple perihelion passages \citep[e.g.,][]{kuhrt84, groussin07,
steckloff16}.  Recent work has emphasized the importance of realistic
material properties, including porosity, tensile strength, and
compositional heterogeneity \citep{attree18, groussin19}.  However,
the case of pristine interstellar comets experiencing first-time solar
approaches requires specialized modeling incorporating high abundances
of supervolatile species (CO, CO$_2$, CH$_4$) that sublimate at much
lower temperatures than water ice and may drive dramatically different
evolutionary pathways.

In this work, we present an advanced numerical framework designed to
explore the fragmentation parameter space for pristine interstellar
comets.  We conduct detailed simulations spanning perihelion distances
from 0.25 to 1.5~AU and tensile strengths from 50 to 500~Pa to identify
the conditions under which catastrophic disruption occurs versus
structural survival.  Our primary focus is a pristine first-timer
scenario representing a comet with high supervolatile content
experiencing thermal shock at 1.0~AU perihelion, which produces dramatic
fragmentation and serves as a limiting case for comparison with
observations.  By comparing our simulation results with the observed
survival of 3I/ATLAS at 1.36~AU, we identify model-dependent
transitions within the adopted parameter space and discuss the
scenarios for material properties and composition with which the
observations are consistent.

Our approach incorporates multiple physical processes operating
simultaneously: (1) a continuous hyperbolic heliocentric trajectory
covering both the inbound and outbound legs of the encounter; (2)
one-dimensional heat conduction into an initially interstellar-cold
(30~K) nucleus; (3) energy-balanced, multi-species volatile sublimation
with retreating subsurface fronts; (4) dust liberation, lifting, and
lag-mantle growth, erosion, and gas sealing; (5) a subsurface
gas-pressure failure criterion against tensile strength, so that
fragmentation---and hence the number of fragments---is an emergent
outcome rather than an input; and (6) N-body dynamics, asymmetric
outgassing torques, and bouncing collisions among the fragments.  This integrated framework enables us to trace the complete
thermal and dynamical evolution from cold interstellar conditions through
perihelion passage and post-disruption fragment dispersal.

Our results provide indicative expectations for the observable
signatures of fragmenting versus surviving interstellar comets,
identify model-dependent critical tensile strengths and compositional
parameters separating these regimes within the adopted parameter
space, and offer theoretical context for interpreting current and
future observations of interstellar visitors.  We stress at the outset
that the present framework is intended as an exploratory numerical
study: whilst its individual physical components are anchored to
laboratory data and to measurements of Solar System comets
(Section~2.10), the integrated pipeline has not yet been benchmarked
end-to-end against a well-observed comet, and its quantitative outputs
should be read accordingly.

\section{Numerical Methods}
\label{sec:methods}

The model described here is a complete rewrite of an earlier pipeline,
undertaken after an audit showed that the
earlier code's fragment count reproduced its rubble-pile input, that its
production rates were dominated by an ad-hoc reservoir cap, and that its
cold-start and outbound-leg behavior were not exercised.  The rewritten
model is deliberately compact; every prescription and parameter is
stated below, and the source code is available from the author upon request.

\subsection{Orbit}
The nucleus follows an exact hyperbolic conic with perihelion $q$ and
hyperbolic excess speed $v_\infty=32$~km\,s$^{-1}$, representative of
the observed interstellar population (26--58~km\,s$^{-1}$ for
1I--3I; \citealt{meech17, guzik20, seligman25}).  The trajectory is
integrated in true anomaly from $r_h=3$~AU inbound, through perihelion,
to $3$~AU outbound, so inbound/outbound asymmetries are genuine model
output.

\subsection{Nucleus and composition}
The nucleus starts as a \emph{single} body with $M_0=2\times10^{12}$~kg
and bulk density $400$~kg\,m$^{-3}$ ($R_0\approx1060$~m), half
refractory dust and half ice by mass (dust/ice $=1$, an adopted value).  The ice is 35\% CO, 30\% CO$_2$, 15\% CH$_4$ and
20\% H$_2$O by mass---fractions of the \emph{icy component}---consistent
with cold ($<$50~K) outer-disc condensation \citep{oberg11,
pontoppidan14} and with the $<$30~K formation temperatures inferred for
3I/ATLAS \citep{roth26, opitom26, salazar26, cordiner26}.  The entire
interior is initialized at the interstellar temperature of 30~K.

\subsection{Thermal model}
Each fragment carries a one-dimensional slab (surface to 6~m depth, 60
geometrically spaced nodes, $\approx$0.6~mm at the surface) solved with
an implicit Crank--Nicolson scheme; the deep boundary is zero-flux.
Constant properties $k_{\rm ice}=0.05$~W\,m$^{-1}$K$^{-1}$,
$c_p=1000$~J\,kg$^{-1}$K$^{-1}$, $\rho=400$~kg\,m$^{-3}$ are adopted, within the ranges reviewed by \citet{groussin19}; a dust mantle of thickness $d_m$ adds a
conductive lid with $k_{\rm mantle}=0.01$~W\,m$^{-1}$K$^{-1}$, combined
in series with the surface cell.  The surface temperature is obtained
each step from the rotation-averaged energy balance
\begin{equation}
(1-A)\frac{L_\odot}{16\pi r_h^2} \;=\; \epsilon\sigma T_s^4
 \;+\; \sum_{i\,\in\,{\rm surf}} L_i Z_i(T_s)
 \;+\; k_{\rm surf}\,\frac{T_s-T_1}{\Delta z},
\label{eq:balance}
\end{equation}
with a low, comet-like albedo $A=0.04$ (cf.\ \citealt{fornasier15}), $\epsilon=0.95$, solved by bisection; the sum runs over
species whose sublimation front is still at the surface.  Sublimation of
\emph{buried} fronts is drawn from the conducted flux (the last term),
so the global energy budget is closed by construction.

\subsection{Sublimation and volatile fronts}
Each species $i$ has vapor pressure $P_i=A_i\exp(-B_i/T)$ with
$B_i=L_i\mu_i/R$ fixed by its latent heat and $A_i$ anchored at the
triple point \citep{fray09}, and free-sublimation flux given by the
Hertz--Knudsen relation $Z_i = P_i\sqrt{\mu_i/2\pi R T}$.  A front at
the surface sublimates at $Z_i(T_s)$ weighted by its ice fraction; a
buried front at depth $z_i$ sublimates at the smaller of its kinetic
rate at the local temperature and the share of the conducted power
reaching it (allocated in order of volatility), further throttled by a
percolation factor $\exp(-z_i/4z_{\rm perm})$ with
$z_{\rm perm}=0.5$~m.  Fronts recede as their local ice column is
consumed.

\subsection{Dust: lifting, mantle growth, erosion, and sealing}
Sublimation liberates dust at the dust/ice ratio.  Grains follow an
MRN-like size distribution ($dn/da\propto a^{-3.5}$, $a\le a_{\rm
top}=0.1$~m, $\rho_{\rm grain}=1500$~kg\,m$^{-3}$); the maximum liftable
size follows from gas drag against the fragment's gravity, and the
unliftable mass fraction accumulates as a lag mantle of density
$450$~kg\,m$^{-3}$.  When outgassing is vigorous enough to lift the
whole distribution, it also erodes an existing mantle at up to the dust
share of the gas flux.  The mantle seals the interior: the effective
confined pressure at a front is $P_{\rm eff}=f_{\rm seal}\,P_i(T_i)$
with $f_{\rm seal}=1-\exp(-d_m/d_{\rm seal})$, $d_{\rm seal}=5$~cm
(fronts buried below 2~m self-confine similarly, but this depth is not
reached in the runs).  This Darcy-like sealing is the model's central
parameterization and is discussed in Section~\ref{sec:limitations}.

\subsection{Failure criterion and splitting}
A fragment fails when $\max_i P_{\rm eff} > \sigma_t$, or when it spins
past the fission limit $\omega^2 > 4\pi G\rho/3 +
\sigma_t/(\rho R^2)$ \citep{scheeres10} under asymmetric outgassing
torques ($f_{\rm asym}=0.2$, $v_{\rm gas}=500$~m\,s$^{-1}$).  Failure
splits the body in two (child mass fraction uniform in 0.25--0.45) with
separation speed $v=\sqrt{2P_{\rm eff}/\rho}$; the fresh faces vent for
$2\times10^5$~s and the mantle is shed, after which mantling and
sealing can recur.  Fragments below $10^{10}$~kg vent instead of
splitting, and a hard cap of 64 tracked bodies guards against runaway.  Because
splitting requires re-mantling, and smaller fragments (weaker gravity)
cannot retain a mantle against gas drag, the cascade self-limits
pre-perihelion (a 53-body plateau lasting $\sim$40 days in the primary
run); the perihelion environment re-triggers it, and the primary run
reaches the cap there, so final counts of 64 are lower bounds set by
the tracking resolution ($M_{\rm min}$) and cap, not converged
physical multiplicities.

\subsection{Fragment dynamics}
Fragments interact through point-mass gravity and a stochastic
asymmetric outgassing thrust; contacts are resolved as inelastic
bounces (restitution 0.3).

\subsection{Mass bookkeeping}
Nucleus mass decreases only through escaping gas and lifted dust, both
of which are accumulated separately, so that
$M_0-M_{\rm final}=M_{\rm gas}+M_{\rm dust}$ holds identically; the
audited residual in the primary run is 0.03~kg out of
$2\times10^{12}$~kg.  Depletion percentages quoted below are fractions
of each species' initial ice reservoir; total mass loss is a fraction
of $M_0$.

\subsection{Parameter space}
We run $q\in\{0.25, 0.8, 1.0, 1.5\}$~AU $\times$
$\sigma_t\in\{50,100,200,500\}$~Pa---spanning and exceeding 67P estimates of $<$150~Pa at 5--30~m scales \citep{groussin15} and a few Pa from overhangs \citep{attree18}---for the
pristine ice composition, plus moderately processed (15\% CO, 20\%
CO$_2$, 10\% CH$_4$, 55\% H$_2$O) and depleted, 67P-like (5\% CO, 10\%
CO$_2$, 5\% CH$_4$, 80\% H$_2$O) classes at $q=1$~AU,
$\sigma_t=100$~Pa.  The primary case ($q=1$~AU, $\sigma_t=100$~Pa) is
run at full resolution; the grid uses a coarser grid/timestep whose
effect is quantified in Section~\ref{sec:resol}.

\subsection{Model Scope and Validation Status}
\label{sec:validation}
We explicitly position this framework as exploratory.  Its ingredients
are anchored to laboratory and cometary data---latent heats and vapor
pressures \citep{fray09}, strengths and thermal properties from Rosetta-era 67P work \citep{groussin15, attree18, groussin19}---but the
integrated pipeline has not been benchmarked end-to-end against a
well-observed comet, and the sealing and splitting prescriptions are
parameterizations.  All thresholds below are therefore model-dependent
trends, not calibrated properties of real objects.

\section{Results}

\subsection{Primary case: emergent disaggregation at $q=1$~AU}
\begin{figure*}
\centering
\includegraphics[width=\textwidth]{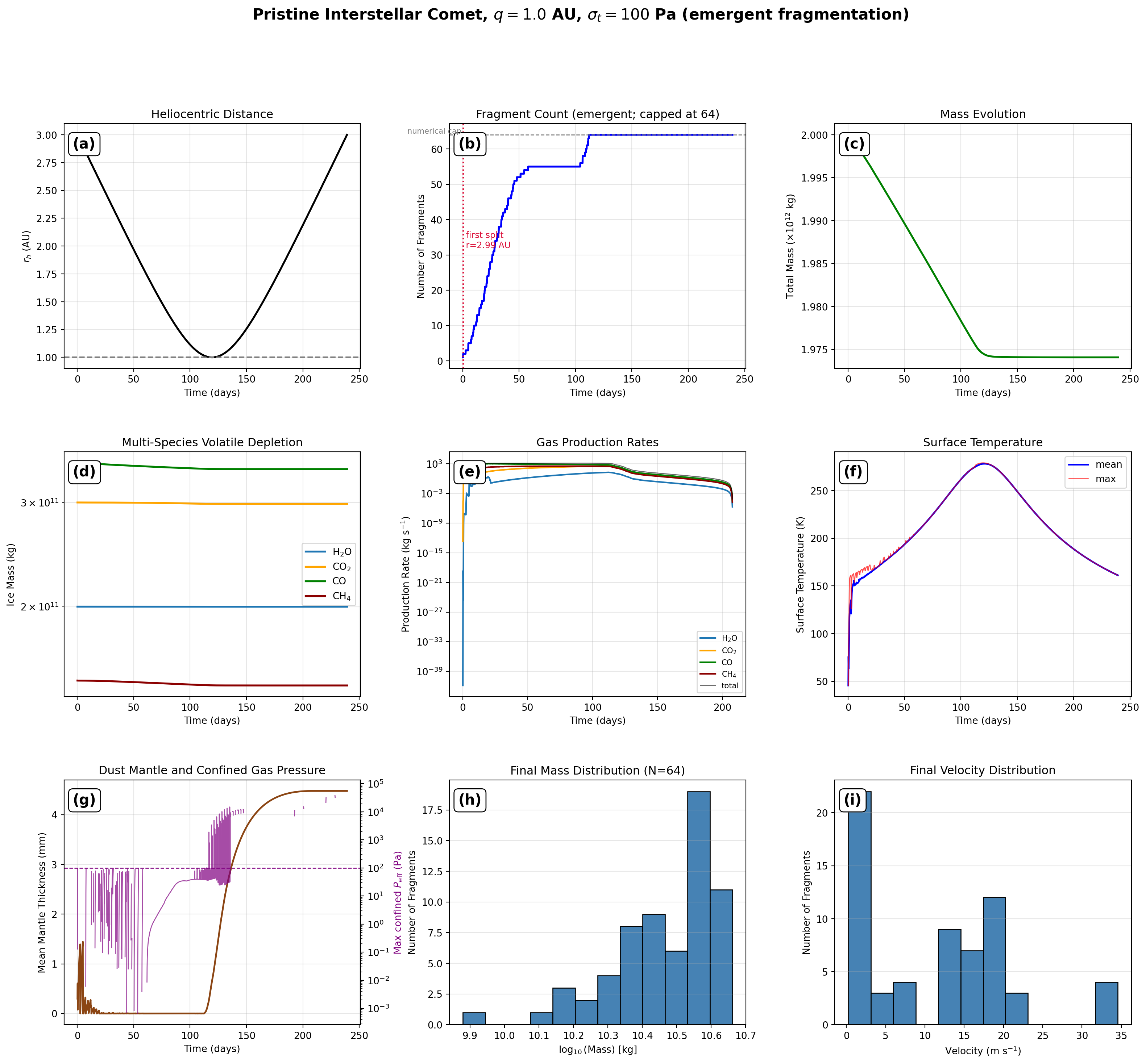}
\caption{Primary simulation ($q=1$~AU, $\sigma_t=100$~Pa, pristine
composition).  (a) Heliocentric distance; (b) emergent fragment count,
with the first splitting marked; (c) total mass; (d) per-species ice
reservoirs; (e) gas production rates (note the physically bounded,
energy-limited scale); (f) mean and maximum surface temperature; (g)
mean mantle thickness and maximum confined gas pressure against
$\sigma_t$ (dashed); (h) final fragment mass distribution; (i) final
velocity distribution relative to the system barycentre.}
\label{fig:primary}
\end{figure*}

Figure~\ref{fig:primary} summarizes the primary run.  Sublimation of CO
begins essentially at the 3~AU start; the resulting sub-lifting-threshold
dust flux deposits a millimeter-scale lag mantle, whose partial sealing
of the warm CO front raises the confined pressure past
$\sigma_t=100$~Pa within the first day, at $r_h=$~2.99~AU---the
onset therefore lies beyond our 3~AU starting distance, consistent
with CO-driven activity observed in comets far outside the water zone
\citep{womack97, jewitt21}.  A cascade of 63 splittings
(62 gas-pressure, 1 rotational-fission events)
follows, self-limiting at 53 bodies for $\sim$40 days pre-perihelion
once fragments are small enough that gas drag prevents mantle
retention; the perihelion thermal environment re-triggers splitting and
the run reaches the 64-body tracking cap, so the final count is a
lower bound of 64.  Total mass loss is 1.3\%
($2.6\times10^{10}$~kg), all of it escaped gas and lifted dust.  The tracking cap is reached at perihelion, so the outbound leg is
necessarily split-free in this run (0 post-perihelion
events); whether the model's mantle-resealing channel would produce
delayed, post-perihelion splittings---as observed for 2I/Borisov
\citep{jewitt20}---is censored by the cap and left to future runs
with a higher tracking limit.

\subsection{Volatile depletion and production rates}
Because sublimation is energy-limited, reservoir depletions remain
modest over a single passage: 2\% (CO), 2\% (CH$_4$),
1\% (CO$_2$) and $<$0.1\% (H$_2$O)---in sharp contrast to
reservoir-exhausting rates, the fragments emerge volatile-rich; the total gas production peaks at
$\approx1.5\times10^{3}$~kg\,s$^{-1}$, an energy-limited value (compare the
$\sim$10$^{2}$--10$^{3}$~kg\,s$^{-1}$ scale of very active Solar System
comets), and the maximum surface temperature reaches 279~K near
perihelion ($t\approx$~120~d).

\subsection{Fragment properties}
Final fragment masses span $7.6\times10^{9}$--$4.6\times10^{10}$~kg (radii
165--301~m at 400~kg\,m$^{-3}$); separation speeds have
median 13.1~m\,s$^{-1}$ (maximum 34.6~m\,s$^{-1}$), set by
$\sqrt{2P_{\rm eff}/\rho}$ at failure and subsequent mutual dynamics.
Outgassing torques spin the fragments up from the initial 10~h to
final periods of 0.6--1.2~h, i.e.\ to near the
size-dependent fission limit, which regulates them: 1 of the
63 splittings are rotational-fission events.  No bouncing
contacts occur after the initial separations (0 collisions).

\subsection{Thermal hysteresis}
\label{sec:hyst}
\begin{figure}
\centering
\includegraphics[width=0.9\columnwidth]{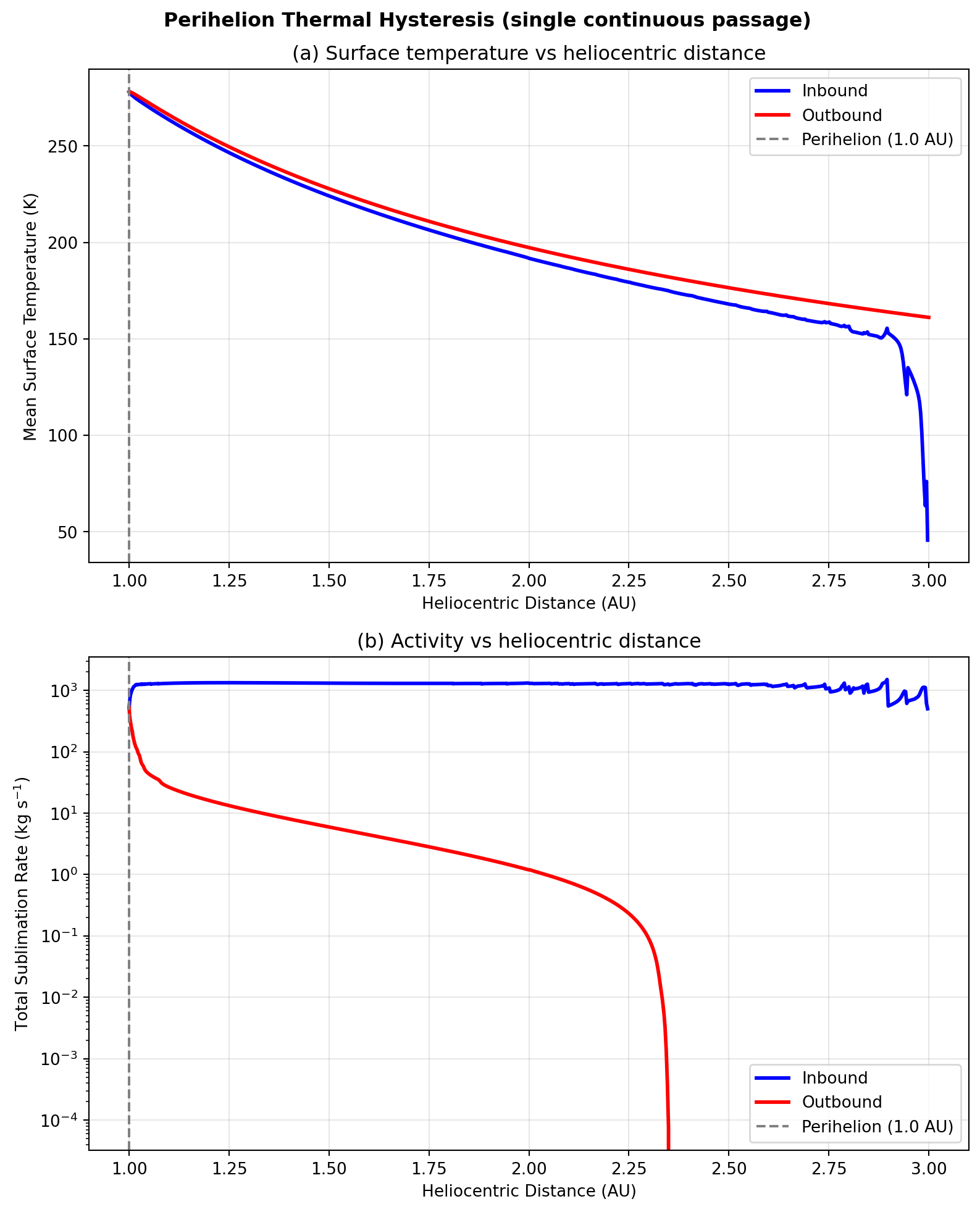}
\caption{Inbound (blue) versus outbound (red) mean surface temperature
(a) and total sublimation rate (b) along the single continuous passage
of the primary run.  Albedo is constant throughout; the asymmetry is
carried by mantling and volatile depletion.}
\label{fig:hyst}
\end{figure}

Figure~\ref{fig:hyst} shows the genuine inbound--outbound asymmetry of
the continuous passage: at $r_h=2$~AU the mean surface temperature is nearly symmetric
(192~K inbound, 197~K outbound---the dry surface simply
tracks radiative equilibrium), while the total sublimation rate is a
factor $\approx$1095.5 higher inbound: the hysteresis lives
almost entirely in the activity, carried by near-surface volatile
depletion and mantling.  The asymmetry is
driven by the mantle--depletion state, not by any albedo change ($A$ is
constant), and its sense---suppressed outbound activity---is qualitatively similar to the faster outbound fading of 3I/ATLAS in post-perihelion Hubble continuum photometry \citep{hui26}, although increased post-perihelion water production has also been reported \citep{cordiner26}, so the observational picture is mixed and may indicate processes outside this model (e.g., mantle erosion exposing fresh ice, or
subsurface water reached only after perihelion; \citealt{prialnik04}).

\subsection{Parameter space}
\label{sec:grid}
\begin{figure*}
\centering
\includegraphics[width=\textwidth]{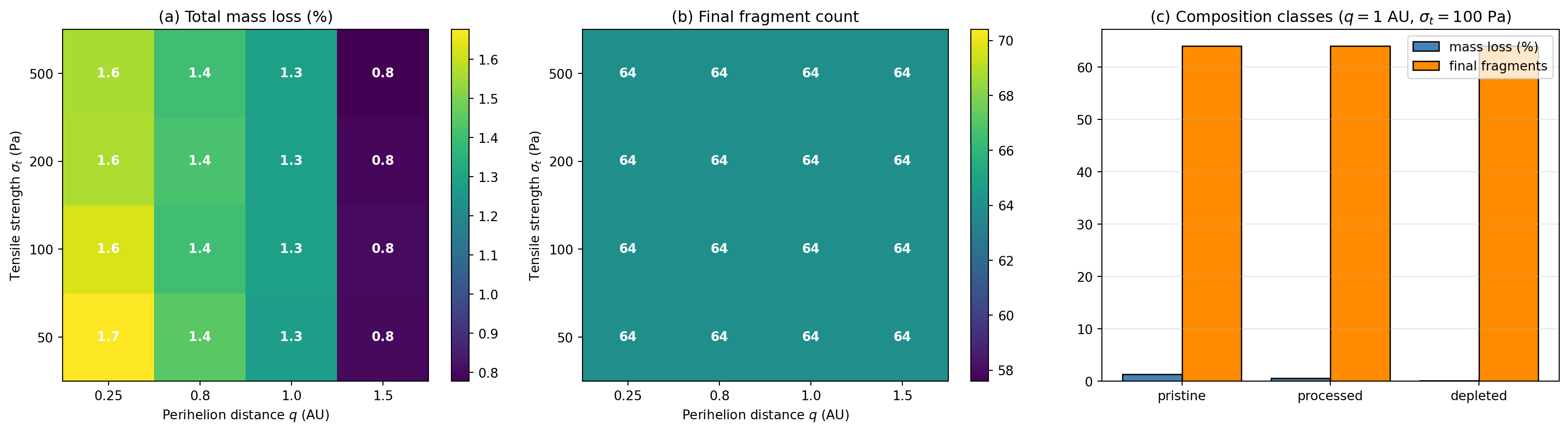}
\caption{(a) Total mass loss and (b) final fragment count across the
perihelion-distance/tensile-strength grid (pristine composition;
reduced resolution, Section~\ref{sec:resol}); (c) composition classes
at $q=1$~AU, $\sigma_t=100$~Pa.}
\label{fig:grid}
\end{figure*}

Figure~\ref{fig:grid} maps total mass loss and final fragment count
over the $q$--$\sigma_t$ grid.  Mass loss spans
0.8--1.7\%, declining gently with perihelion
distance, and every grid run likewise disaggregates to the 64-body tracking cap.  Within this model, no combination leaves the nucleus intact.  The
weak dependence on $\sigma_t$ reflects the exponential temperature
dependence of the confined vapor pressure: once a partially sealed CO
front warms past $\sim$50~K its pressure rises orders of magnitude in a
few kelvin, so a factor-of-ten change in $\sigma_t$ shifts the failure
epoch only modestly.  Perihelion distance and, above all, composition
control the outcome: the processed and depleted classes at $q=1$~AU,
$\sigma_t=100$~Pa lose 0.6\% and 0.1\% of their mass,
versus 1.3\% for pristine ices---a factor
$\approx$10.7 between the pristine and depleted classes.  All
three classes nonetheless disaggregate to the tracking cap: in this
model the failure trigger depends on the \emph{temperature} of a
sealed front, not on the reservoir size, so even trace supervolatiles
suffice to trip it, while abundance controls the mass-loss budget.
This is the model's strongest---and most suspect---prediction; see
Section~\ref{sec:3i}.

\subsection{Resolution and stochastic sensitivity}
\label{sec:resol}
Re-running the primary case at the grid resolution (40 nodes, doubled
timestep) gives 1.3\% mass loss (versus 1.3\% at full
resolution) with an essentially identical splitting history, and three
random seeds at grid resolution span 1.3--1.3\% in mass loss, with
every realization reaching the tracking cap.  Mass loss is thus converged and seed-robust at the quoted
precision.  The final fragment count, by contrast, saturates the
tracking cap in all realizations and is bounded below by construction;
it should be read as ``at least several tens above
$M_{\rm min}=10^{10}$~kg,'' not as a converged multiplicity.

\section{Discussion}

\subsection{Scenarios consistent with the survival of 3I/ATLAS}
\label{sec:3i}
Within this model, no explored configuration survives intact:
no combination leaves the nucleus intact, and even at $\sigma_t=500$~Pa and $q=1.5$~AU---bracketing
3I/ATLAS's $q=1.36$~AU---the pristine nucleus fragments
(0.8\% mass loss).  Because the depleted, 67P-like class also
disaggregates while real Jupiter-family comets of that composition
survive repeated perihelia, we do not read the model's universal
fragmentation as a statement about real objects; rather, the sealing
prescription (Section~\ref{sec:limitations}) is evidently too
aggressive, and absolute survival predictions are outside what this
calibration can deliver.  What the grid does support for 3I/ATLAS ($q=1.36$~AU; \citealt{seligman25}) is a ranking: outcomes are controlled by
supervolatile content and perihelion distance, and are nearly
independent of tensile strength, so the intact survival of 3I is far
more informative about its (non-pristine) surviving ice inventory or
its venting efficiency than about its strength.  If galactic-transit processing devolatilized the outer layers of 3I---a possibility not modeled here---then the isotopically
inferred cold formation of 3I \citep{roth26, opitom26, salazar26,
cordiner26} need not imply CO-rich surviving ice today.

\subsection{Comparison with 2I/Borisov}
The model contains a mechanism naturally suited to delayed splittings:
as activity declines, dust lifting weakens, mantles reform, and
resealing can rebuild confined pressure.  In the present runs this
channel is censored---the tracking cap is saturated by perihelion, so
no post-perihelion splittings can occur numerically---and we therefore
advance it only as a candidate mechanism for 2I/Borisov's minor
fragmentation $\sim$3 months post-perihelion, complementing the rotational-bursting interpretation of \citet{jewitt20}, to be tested with an
uncapped configuration.

\subsection{Dominant controls}
Ranking the model's sensitivities: (1) ice composition (supervolatile
fraction) dominates both mass loss and fragmentation; (2) perihelion
distance sets the insolation history; (3) the sealing parameters
($d_{\rm seal}$, mantle retention) control whether confined pressure
can build at all; (4) tensile strength enters only logarithmically
through the failure temperature; (5) rotation and collisions are
subdominant here.  A global sensitivity analysis over this vector
remains future work.

\subsection{Limitations and future directions}
\label{sec:limitations}
The central caveats are: (1) the Darcy-like sealing law
($f_{\rm seal}$, $d_{\rm seal}$) is a parameterization, not a solved
gas-diffusion problem \citep{skorov11, davidsson22}; (2) splitting is
binary with a prescribed mass-ratio distribution, without fracture mechanics; (3) thermal transport is 1-D per
fragment with constant properties, omitting lateral conduction,
amorphous-ice crystallization energy \citep{prialnik87, mumma11}, and
layered heterogeneity; (4) a single passage is simulated, from and to
3~AU, although CO sublimates far beyond that distance
\citep{womack97, jewitt21}, so pristine reservoirs may be partly
processed before 3~AU and our depletions are upper limits in that
sense; (5) the final fragment count saturates the 64-body tracking cap and is
bounded by $M_{\rm min}$, so only its order of magnitude is
meaningful (Section~\ref{sec:resol}); and (6) the sealing law makes
the failure trigger independent of volatile abundance, which is why
even depleted ices fragment here while real comets of that composition
survive---the single clearest target for calibration.  Benchmarking against 67P's measured
activity and against 2I/Borisov, and replacing the seal law with a
porous-flow solution, are the immediate next steps.

\section{Conclusions}
We rebuilt our exploratory framework so that fragmentation is an
emergent outcome of a closed energy and mass budget.  A pristine, CO-rich interstellar comet of comet-like
strength begins disaggregating essentially at our 3~AU starting
distance in the primary case, ends at the tracking cap ($\geq$64 fragments) with 1.3\%
mass loss, and---across $q=0.25$--$1.5$~AU and
$\sigma_t=50$--$500$~Pa---no combination leaves the nucleus intact.  Composition and perihelion distance, not strength, are the decisive
parameters in this model; the calibration over-fragments (even
67P-like ices disaggregate, unlike real Jupiter-family comets), so we
advance the composition ranking, the $\sigma_t$-insensitivity, the
percent-level energy-limited mass loss, and the beyond-3-AU fragmentation onset as the robust results (the
mantle-resealing route to delayed, 2I/Borisov-like splittings is
cap-censored here and awaits uncapped runs), and defer absolute
survival statements to a benchmarked successor model.  All thresholds are
model-dependent trends of a parameterized, un-benchmarked framework;
the released code, closed budgets, and stated prescriptions are
intended to make every number in this paper reproducible and
falsifiable.  The Vera C.~Rubin Observatory's forthcoming interstellar
discoveries \citep{engelhardt17} will test the model's central, falsifiable tendency: CO-rich
first-passage comets at $q\lesssim1.5$~AU should fragment more
readily, and lose relatively more mass, than volatile-depleted ones at
the same perihelion.

\section*{Declaration on the Use of Generative AI}

The author used a large language model (LLM) solely for minor language editing and improvements 
to clarity and presentation of the manuscript. The underlying Python pipeline, methodology, analysis, 
and software implementation were conceived and developed entirely by the author without LLM assistance.

\section*{Data and Code Availability}
The complete rewritten pipeline (\texttt{comet\_fragmentation\_v2.py}),
run drivers, and all run outputs used in the figures is available from the author upon request.

\section*{Acknowledgements}
B.K. acknowledges helpful discussions with colleagues on comet
fragmentation physics and interstellar object dynamics.

\end{document}